\documentclass[aps,pre,reprint,showpacs,showkeys,amsmath]{revtex4-1}
\usepackage{hyperref}
\usepackage{graphicx}

\begin{document}

\title{Externally-driven collisions of domain walls in bistable systems near criticality}
\author{Andrzej Janutka}
\email{Andrzej.Janutka@pwr.wroc.pl}
\affiliation{Institute of Physics, Wroclaw University of Technology,
Wybrze\.ze Wyspia\'nskiego 27, 50-370 Wroc{\l}aw, Poland}

\begin{abstract}
Multi-domain solutions to the time-dependent Ginzburg-Landau equation in presence of an external field 
 are analyzed using the Hirota bilinearization method. Domain-wall collisions are studied in detail considering
 different regimes of the critical parameter. I show the dynamics of the Ising and Bloch domain walls of the Ginzburg-Landau
 equation in the bistable regime to be similar to that of the Landau-Lifshitz domain walls. Domain-wall
 reflections lead to the appearance of bubble and pattern structures. Above the Bloch-Ising transition point, 
 spatial structures are determined by the collisions of fronts propagating into an unstable state. 
 Mutual annihilation of such fronts is described.   
\end{abstract}


\maketitle
\newpage

\section{Introduction}

Localized and labyrinth structures in excitable media are considered to be used as carriers of binary information for specific 
 applications \cite{cou04,mik06}. Such structures appear in monostable and bistable regimes of magnetic, electrical, chemical,
 and optical systems. In particular, different realizations of memory and logic elements with bubble-forming bistable media 
 seem to be technologically promising because of a high stability of the bubble structures. 
 Prototype bubble-forming systems are 2D ferromagnets including ferrofluids \cite{del80,rom75}, and 2D ferroelectrics
 including electrically active liquid crystals \cite{pir93,kor04}. In critical parameter ranges, these systems create
 pattern structures (lamellae, labyrinths) which are not periodic \cite{del80,wu07}. To be precise, I mention that 
 experimental observation of bubbles and lamellar patterns in solid ferroelectric films does not manage (unlike
 in liquid crystals where electro-elastic coupling is weaker) \cite{sco07}, although a 2D bubble like structure
 has been seen in a system of a finite-size geometry \cite{gru08}. Unlike patterns in oscillatory critical systems,
 the above structures are built of domain walls (DWs) connecting stable states of opposite orientation of an order 
 parameter. Similar patterns and bubbles are observed in excitable (reaction-diffusion) chemical systems below
 the critical (bifurcation) point \cite{lee93,lee95}. It is hoped that difficulties of early 
 magnetic and dipolar bubble technologies reviewed in \cite{ode86,sco89} can, nowadays, be overcome thanks to the advance 
 of miniaturization. One observes increased interest in studies of DW complexes with relevance to data storage
 and processing devices which are based on ferromagnetic and ferroelectric nanoelements. In particular, nanowire structures
 are under intense investigation because of the possibility of the most dense information packing \cite{par08,sch07}.
 The importance of critical effects in nanosystems grows with decreasing in their diameters because of a significant
 decrease of the critical (Neel) temperature \cite{zen02,str02}. 

Generic critical properties of the magnetic and electric media are described with the time-dependent Ginzburg-Landau
 equation which enables qualitative analysis of DW behaviors in all bistable systems. The DW solutions to the
 1D Ginzburg-Landau equation are of Ising (named Neel DWs also) or Bloch type. Under external 
 stimulation the DWs move and their collision properties determine morphology of resulting pattern structures.
 In the present paper, external-field-induced DW collisions in 1D Ginzburg-Landau systems are studied with reference 
 to bubble formation. I show similarities 
 of the process to the bubble formation in ferromagnetic (or ferroelectric) wire far from the criticality (a system
 described with the Landau-Lifshitz-Gilbert equation), referring to my previous study of the problem \cite{jan10}
 as I throughout the text \cite{jan10}. The creation of bubbles and DW patterns is described with connection to the property 
 of elastic reflection of DWs observed in excitable media. However, I emphasize that the present study
 is not related to widely investigated pulses in monostable regimes of excitable systems \cite{bar92,cro09}.
 
In sections II and III, the field-induced collisions of DWs and phase-fronts, respectively, are studied. In section IV,
 the relevance of predictions on DW and front collisions to pattern formation is discussed.  
 
\section{Domain wall collisions} 
 
Let us consider 1D Ginzburg-Landau equation for the ferromagnetic (ferroelectric) wire in a (longitudinal) external 
 field directed along the spontaneous magnetization (polarization)
\begin{eqnarray}
\alpha\frac{\partial m}{\partial t}=J\frac{\partial^{2}m}{\partial x^{2}}
+\beta_{1}m+\beta_{2}m^{*}-\mu|m|^{2}m+\gamma H.
\label{GL}
\end{eqnarray}
Here $m$ denotes a complex order parameter (a two-component magnetization or polarization), $\beta_{1}$ determines 
 the distance from the criticality ($\beta_{1}=-\beta_{2}$ at the phase-transition point), $H$ denotes an external (magnetic 
 or electric) field intensity (it can take positive or negative real values). Basing on arguments related to the time-reversal
 symmetry, I show that the described with (\ref{GL}) DWs have to reflect upon their collision in the parameter region where
 they form magnetic bubbles. Under the assumption $\gamma H\ll(\beta_{1}+\beta_{2})^{3/2}/\mu^{1/2}$, the domains of the stable phase
 of the system correspond to $m\approx\pm\sqrt{\beta_{1}+\beta_{2}}/\sqrt{\mu}$, and the single-DW solutions to (\ref{GL})
 take the form 
\begin{eqnarray}
m&=&\sqrt{\frac{\beta_{1}+\beta_{2}}{\mu}}\frac{1-{\rm e}^{2\eta}}{1+{\rm e}^{2\eta}}
+{\rm i}\theta2\sqrt{\frac{\beta_{1}-3\beta_{2}}{\mu}}\frac{{\rm e}^{\eta}}{1+{\rm e}^{2\eta}},
\nonumber\\
\eta&=&k(x-x_{0})-\frac{\sqrt{\mu}\gamma Ht}{\sqrt{\beta_{1}+\beta_{2}}\alpha},
\label{DW}
\end{eqnarray}
where
\begin{subequations}
\begin{eqnarray}
|k|=\sqrt{\frac{2\beta_{2}}{J}}\hspace*{2em}{\rm for}\hspace*{2em}\theta=\pm1,
\label{BlochDW}
\end{eqnarray}
and
\begin{eqnarray}
|k|=\sqrt{\frac{\beta_{1}+\beta_{2}}{2J}}\hspace*{2em}{\rm for}\hspace*{2em}\theta=0.
\label{IsingDW}
\end{eqnarray}
\end{subequations}
These equations describe the Bloch DW and the Ising DW, respectively. Only in the limiting case $H=0$, the above solution
 is strict while, for $H\neq 0$, it neglects a small field-induced deformation of the DW profile (an space asymmetry arising
 with respect to the DW center; details are given in Appendix A).
 The DWs move under the condition $H\neq 0$ in the direction dependent on the sign of $k$. 
 It is seen from (\ref{DW}),(\ref{BlochDW}) that the Bloch DWs can exist only in the regime 
 $\beta_{1}>3\beta_{2}$; therefore, at $\beta_{1}\to3\beta_{2}^{+}$, the Bloch DWs transform into the Ising DWs
 which are the only known stationary front solutions in the regime $-\beta_{2}<\beta_{1}<3\beta_{2}$ \cite{laj79}.
 Stationary two-DW solutions to (\ref{GL}) are given in \cite{bar05}. In absence of the external
 field $H=0$, the double-wall solutions describe localized static bubbles built of pairs of different type
 (Ising and Bloch) DWs. The size of the bubble can be changed by the application of a field $H\neq 0$ because of 
 inducing the motion of its fronts (DWs) and, eventually, their collisions. The collision result (DW reflection or annihilation)
 determines final spatial structure created by DWs, similar to whose of 1D ferromagnets far from the critical regime (I).
 In the present work, I omit the discussion of the formation of bubbles built of two Bloch DWs or two Ising DWs mentioning 
 previous studies of Refs \cite{mal94}. Such objects are unstable for $H=0$ (initially stationary DWs repeal or attract 
 each other), thus, they are less important for potential applications than Ising-Bloch bubbles (although, 
 the stability of the Ising-Bloch bubbles against temporal external perturbations is a separate question discussed 
 in \cite{bar07a}).
   
Obviously, Eq. (\ref{GL}) is not time-reversal invariant, while the reversibility of a dynamical
 system is a crucial property in terms of the stability of any localized structure (e.g. a bubble) \cite{cou04,pom86}. 
 Due to this irreversibility, solitary-wave solutions to (\ref{GL}) are, in the general case, relevant only to the limit 
 of large positive values of time ($t\to\infty$), while they are irrelevant to the opposite limit ($t\to-\infty$), 
 in particular, because of infinite growth of the energy with $t\to-\infty$ which is indicated by nonzero values 
 of the dissipative function (I). With relevance to studying the DW collisions, and thus to investigating the stability of DW complexes,
 I analyze the dynamics of the bubble in the limit of large negative values of time (in this limit, the colliding DWs 
 are noninteracting objects), modifying the evolution equation (\ref{GL}) by reversing the arrow of time. The inverse-evolution
 equation takes the form  
\begin{eqnarray}
-\alpha\frac{\partial\tilde{m}}{\partial t}&=&J\frac{\partial^{2}\tilde{m}}{\partial x^{2}}
+\beta_{1}\tilde{m}+\beta_{2}\tilde{m}^{*}-\mu|\tilde{m}|^{2}\tilde{m}-\gamma H.
\label{tilde-GL}
\end{eqnarray}
Let me mention an analogy to the necessity of doubling the number of degrees of freedom when formulating 
 the dynamics of dissipative (classical or quantum) systems within formalisms relevant to the whole length 
 of the time axis \cite{bat31,kel65}, (see also I), in particular, the dynamics of essentially dissipative 
 (reaction-diffusion) systems whose excitations are overdamped \cite{doi76}. 
%I have doubled the number 
% of freedom degrees in (I) when studying the DW motion far from the criticality. I base my extension of the 
% Ginzburg-Landau equation (as well as the previous extension of the Landau-Lifshitz-Gilbert equation of I) 
% on the claim that far enough from the critical point, where the system is in 
% the thermodynamic equilibrium, its dynamical equations should be symmetric to the time reversal.
% By analogy to thermodynamics, where any reversible process is an equilibrium one (a quasi-static one),
% the full equation of motion describing phenomena in an equilibrium state (e.g. linear waves, DW motion) 
% should be a time-reversal symmetric (reversible) one.    

Applying the Hirota's bilinearization method, following the substitutions $m=g_{1}/f_{1}$, 
 $\tilde{m}=-g_{2}/f_{2}$ in (\ref{GL}) and (\ref{tilde-GL}), where $f_{1}$, $f_{2}$ take real values, 
 one arrives at the secondary equations of motion
\begin{subequations}
\begin{eqnarray}
(-\alpha D_{t}+JD_{x}^{2})g_{1}\cdot f_{1}+(\beta_{1}-\lambda)g_{1}f_{1}&&
\nonumber\\
+\beta_{2}g_{1}^{*}f_{1}+\gamma Hf_{1}^{2}&=&0,
\nonumber\\
JD_{x}^{2}f_{1}\cdot f_{1}+\mu g_{1}g_{1}^{*}-\lambda f_{1}^{2}&=&0,
\label{secondary_GLa}\\
(\alpha D_{t}+JD_{x}^{2})g_{2}\cdot f_{2}+(\beta_{1}-\lambda)g_{2}f_{2}&&
\nonumber\\
+\beta_{2}g_{2}^{*}f_{2}+\gamma Hf_{2}^{2}&=&0,
\nonumber\\
JD_{x}^{2}f_{2}\cdot f_{2}+\mu g_{2}g_{2}^{*}-\lambda f_{2}^{2}&=&0.
\label{secondary_GLb}
\end{eqnarray}
\end{subequations}
Here, 
\begin{eqnarray}
D_{t}^{m}D_{x}^{n}b(x,t)\cdot c(x,t)\equiv
(\partial/\partial t-\partial/\partial t^{'})^{m}
\nonumber\\
\times
(\partial/\partial x-\partial/\partial x^{'})^{n}b(x,t)c(x^{'},t^{'})|_{x=x^{'},t=t^{'}}.
\nonumber
\end{eqnarray}
The above breaking of (\ref{GL}), (\ref{tilde-GL}) into the pairs the secondary equations (\ref{secondary_GLa}), 
 (\ref{secondary_GLb}), respectively, is nonunique but it leads to the equations of the lowest possible order 
 in $g_{1(2)}$, $f_{1(2)}$ (bilinear ones). Upon the inversion of the arrow of time ($t\to-t$), the secondary dynamical
 variables transform following $g_{1(2)}\to-g_{2(1)}$, $f_{1(2)}\to f_{2(1)}$ while (\ref{secondary_GLa}) transform 
 into (\ref{secondary_GLb}) and vice versa. 
 
For $\lambda=\beta_{1}+\beta_{2}$ and $\gamma H\ll(\beta_{1}+\beta_{2})^{3/2}/\mu^{1/2}$, (the field is much weaker then 
 the coercivity value; thus, $m\approx\pm\sqrt{\beta_{1}+\beta_{2}}/\sqrt{\mu}$ inside the domains and a deformation 
 of the DWs induced by $H$ is negligible), following Appendix A, I find a two-DW (bubble) solutions in the form 
\begin{subequations}
\begin{eqnarray}
g_{1}&=&a(1-\nu{\rm e}^{2\eta_{1}}-\nu{\rm e}^{2\eta_{2}}+{\rm e}^{2\eta_{1}+2\eta_{2}})
+{\rm i}b\nu^{1/2}{\rm e}^{\eta_{1}}(1-{\rm e}^{2\eta_{2}}),
\nonumber\\
f_{1}&=&1+\nu{\rm e}^{2\eta_{1}}+\nu{\rm e}^{2\eta_{2}}+{\rm e}^{2\eta_{1}+2\eta_{2}},
\nonumber\\
\eta_{j}&=&k_{j}(x-x_{0j})-\omega_{j}t,
\label{doubleDW_Ia}\\
g_{2}&=&a(1-\nu{\rm e}^{2\tilde{\eta}_{1}}-\nu{\rm e}^{2\tilde{\eta}_{2}}
+{\rm e}^{2\tilde{\eta}_{1}+2\tilde{\eta}_{2}})
+{\rm i}b\nu^{1/2}{\rm e}^{\tilde{\eta}_{1}}(1-{\rm e}^{2\tilde{\eta}_{2}}),
\nonumber\\
f_{2}&=&1+\nu{\rm e}^{2\tilde{\eta}_{1}}+\nu{\rm e}^{2\tilde{\eta}_{2}}+{\rm e}^{2\tilde{\eta}_{1}+2\tilde{\eta}_{2}},
\nonumber\\
\tilde{\eta}_{j}&=&k_{j}(x-x_{0j})+\omega_{j}t,
\label{doubleDW_Ib}
\end{eqnarray}
\end{subequations}
where
\begin{eqnarray}
a=\sqrt{\frac{\beta_{1}+\beta_{2}}{\mu}},\hspace*{3em}b=2\sqrt{\frac{\beta_{1}-3\beta_{2}}{\mu}},
\nonumber\\
k_{1}=\mp\sqrt{\frac{\beta_{1}+\beta_{2}}{2J}},\hspace*{1em}k_{2}=\pm\sqrt{\frac{2\beta_{2}}{J}},\hspace*{1em}
\omega_{1}=\omega_{2}=
\nonumber\\
=\frac{\sqrt{\mu}\gamma H}{\sqrt{\beta_{1}+\beta_{2}}\alpha},\hspace*{1em}
\nu=\frac{\beta_{1}+\beta_{2}-2\sqrt{\beta_{2}(\beta_{1}+\beta_{2})}}{\beta_{1}+\beta_{2}+2\sqrt{\beta_{2}(\beta_{1}+\beta_{2})}}.
\label{doubleDW_II}
\end{eqnarray}
In the limit $H\to 0$, 
 $g_{1}=g_{2}$, $f_{1}=f_{2}$, and both the fields $m$ and $-\tilde{m}$ coincide with the static bubble solution to (\ref{GL}) 
 written in \cite{bar05}.

Studying the DW collision, I analyze the limit $t\to\infty$ of the double DW solutions to (\ref{GL}), the formula (\ref{doubleDW_Ia}),
 and the $t\to-\infty$ limit of the solution to (\ref{tilde-GL}), the formula (\ref{doubleDW_Ib}), in the closest vicinity of the $i$-th 
 DW center, noticing that
\begin{eqnarray}
m\approx m^{(i)}=a\frac{1-{\rm e}^{2\eta_{i}+{\rm ln}(\nu)}}{1+{\rm e}^{2\eta_{i}+{\rm ln}(\nu)}}
+{\rm i}\delta_{i2}b\frac{{\rm e}^{\eta_{i}+{\rm ln}(\nu)/2}}{1+{\rm e}^{2\eta_{i}+{\rm ln}(\nu)}}
\label{initial}
\end{eqnarray}
for $0\approx \eta_{i}+{\rm ln}(\nu)/2\gg\eta_{k}+{\rm ln}(\nu)/2$ and $i\neq k$,
\begin{eqnarray}
\tilde{m}\approx\tilde{m}^{(i)}=-a\frac{1-{\rm e}^{2\tilde{\eta}_{i}+{\rm ln}(\nu)}}{1+{\rm e}^{2\tilde{\eta}_{i}+{\rm ln}(\nu)}}
-{\rm i}\delta_{i2}b\frac{{\rm e}^{\tilde{\eta}_{i}+{\rm ln}(\nu)/2}}{1+{\rm e}^{2\tilde{\eta}_{i}+{\rm ln}(\nu)}}
\end{eqnarray}
for $0\approx\tilde{\eta}_{i}+{\rm ln}(\nu)/2\gg\tilde{\eta}_{k}+{\rm ln}(\nu)/2$ and $i\neq k$. Note that (\ref{initial}) coincides
 with the Ising-DW profile (\ref{DW}),(\ref{IsingDW}) for $i=1$ and with the Bloch-DW profile (\ref{DW}),(\ref{BlochDW}) 
 for $i=2$. In order to establish the magnetization dynamics in the limit of large negative values of time, I invert 
 the motion of $\tilde{m}^{(i)}$ utilizing the property $\tilde{m}^{(i)}(x,0)=-\tilde{m}^{(i)*}(-x+2x_{0i}^{'},0)$, where
 $x_{0i}^{'}=x_{0i}-{\rm ln}(\nu)/(2k_{i})$, and find
\begin{eqnarray}
m\approx-a\frac{1-{\rm e}^{2\eta_{i}+{\rm ln}(\nu)}}{1+{\rm e}^{2\eta_{i}+{\rm ln}(\nu)}}
+{\rm i}\delta_{i2}b\frac{{\rm e}^{\eta_{i}+{\rm ln}(\nu)/2}}{1+{\rm e}^{2\eta_{i}+{\rm ln}(\nu)}}
\label{final}
\end{eqnarray}
for $0\approx \eta_{i}+{\rm ln}(\nu)/2\ll\eta_{k}+{\rm ln}(\nu)/2$ and $i\neq k$.  
 Comparing (\ref{initial}) and (\ref{final}), one sees that the result of the collision is a change 
 of the sign of the real part of $m^{(i)}$. Thus, one can think of the colliding DWs that they pass through each other
 with constant velocity and change their character from the head-to-head into the tail-to-tail structure and vice versa.
 In fact, the result of the DW collision is their elastic reflection accompanied by change of the Bloch DW into Neel DW 
 and vice versa, similarly to the result of field-driven or spontaneous collision of magnetic DWs of the 1D Landau-Lifshitz equation (I). 
 Let me emphasize that taking the $t\to-\infty$ limit of $g_{1}/f_{1}$ at $\eta_{1(2)}+{\rm ln}(\nu)/2\approx 0$ leads to $m$ differing 
 from (\ref{final}) by the sign of its imaginary part, which confirms the irrelevance of (\ref{GL}) in this limit.
 
Above, I analyzed an infinite 1D medium.
 Finite length of the domains leads to additional consequences of the DW collision because of the natural tendency 
 of finite systems toward energy minimization (the energy cannot be defined for infinite systems).
 The preferred direction of the DW motion before the collision (the growth or decrease of the bubbles) 
 corresponds to the decrease of the Zeeman energy while the collision-induced motion in the reversed direction 
 finishes when the decrease of the interaction energy of the (repulsing) DWs equals the Zeeman-energy increase. 
 Thus, I predict the appearance of localized bubbles due to an external-field application in the parameter range
 $\beta_{1}>3\beta_{2}$. Their diameter is determined by the strength of the driving field. 
 
Approaching the Bloch-Ising transition point with $\beta_{1}$ results in a decrease of $\nu$, 
 ($\beta_{1}\to 3\beta_{2}^{+}\Rightarrow\nu\to 0^{+}$). Then, the imaginary parts of $m$, $\tilde{m}$ decrease and,
 eventually, vanish at $\beta_{1}=3\beta_{2}$. Simultaneously, the centers of both the DWs move away from each other 
 due to the shift of the parameters (\ref{doubleDW_II}) of (\ref{doubleDW_Ia}) and (\ref{doubleDW_Ib}), ($\nu\to\nu+\Delta\nu$,
 $k_{j}\to k_{j}+\Delta k_{j}$, $\Delta k_{2}=0$). The shifts of the DW centers are equal to 
 $\Delta x_{0j}={\rm ln}(\nu)/|2k_{j}|-{\rm ln}(\nu+\Delta\nu)/|2k_{j}+2\Delta k_{j}|$. 
 Finally, at $\beta_{1}=3\beta_{2}$, ($\Delta\nu=-\nu$, $k_{j}+\Delta k_{j}=\sqrt{2\beta_{2}/J}$) the centers of 
 both the DWs diverge. A domain pattern is expected to appear due to finite density of the mutually interacting 
 bubbles. Further decrease of $\beta_{1}$ (down to $-\beta_{2}$) results in divergence of the widths of the (Ising) DWs and, 
 thus, in the divergence of the length of the DW interaction. An adiabatic approaching with $\beta_{1}\to-\beta_{2}^{+}$
 is impossible in practice due to the instant widening of the DW. Because of the rapid increase of the overlap of DWs, 
 one anticipates the appearance of domains of an unstable phase in our bistable medium in the regime 
 $-\beta_{2}<\beta_{1}<3\beta_{2}$ (above the Bloch-Ising transition and below the bifurcation point).
 Then, an energy excess is being removed from the system via the propagation of fronts connecting domains
 of a stable phase with domains of an unstable phase (e.g. fronts between ferromagnetic and paramagnetic 
 domains). They always propagate into unstable state and below I call them phase fronts \cite{saa03}. 
% Since the equilibrium disturbance in the domains results
% in breaking the dynamics reversibility, in this regime of the critical parameter, there is no reason 
% for phase-front effects to be described by a time-reversal invariant equation of motion. 
 
\section{Phase-front collisions}

We can gain insight into the pattern formation near the criticality by considering the collision of fronts
 propagating into unstable states. Let us study front solutions to (\ref{GL}) assuming ${\rm Im}(m)=0$ 
 or, alternatively, to a simplified version of (\ref{GL}),  
\begin{eqnarray}
-\alpha\frac{\partial m}{\partial t}+J\frac{\partial^{2}m}{\partial x^{2}}
+\theta m-\mu m^{3}+\gamma H=0,
\label{simplified-GL}
\end{eqnarray}
where $\theta=\beta_{1}+\beta_{2}$, (the real solutions to both the equations are the same). Equation (\ref{simplified-GL}) 
 is relevant to the closest vicinity of the critical point $\theta=0$, where the order parameter 
 takes real values. On the other hand, ${\rm Im}(m)$ vanishes when crossing the Bloch-Ising transition point
 $\beta_{1}=3\beta_{2}$; thus, real solutions are expected to qualitatively describe the phase-front motion 
 for the parameter range $-\beta_{2}<\beta_{1}<3\beta_{2}$. Since $m\to 0$ with $\theta\to 0$, I introduce a renormalized 
 field $m^{'}\equiv m/\sqrt{\theta}$ in order to determine dynamical properties of the phase fronts in the vicinity 
 of the critical point. Substituting $m^{'}$ by $g/f$, one arrives at a trilinear (Hirota) form of (\ref{simplified-GL});
\begin{eqnarray}
f[(-\alpha D_{t}+JD_{x}^{2})g\cdot f+\gamma H/\sqrt{\theta}f^{2}]&&
\nonumber\\
+g\left[(-JD_{x}^{2}+\theta)f\cdot f-\mu\theta g^{2}\right]&=&0.
\label{simplified-GL-II}
\end{eqnarray}
For $H=0$, an exact single-front solution 
\begin{eqnarray}
g=\frac{1}{\sqrt{\mu}}{\rm e}^{k(x-x_{01})+ckt},\hspace*{1em}f=1+{\rm e}^{k(x-x_{01})+ckt},
\end{eqnarray}
here
\begin{eqnarray}
k=\sqrt{\frac{\theta}{2J}},\hspace*{3em}ck=\frac{3\theta}{2\alpha},
\end{eqnarray}
and a double-front solution to (\ref{simplified-GL-II}) 
\begin{eqnarray}
g=\frac{1}{\sqrt{\mu}}({\rm e}^{k(x-x_{01})+ckt}-{\rm e}^{-k(x-x_{02})+ckt}),
\nonumber\\
f=1+{\rm e}^{k(x-x_{01})+ckt}+{\rm e}^{-k(x-x_{02})+ckt}
\label{Nozaki-Bekki}
\end{eqnarray}
were found by Nozaki and Bekki \cite{noz84}. I mention that these authors have considered a generalization of (\ref{simplified-GL})
 exchanging $\alpha$, $J$, $\mu$ into complex coefficients (and allowing complex $m$), however, the real parts of these 
 coefficients were found to determine generic dynamical properties of the phase fronts whereas the imaginary parts 
 are responsible mainly for additional periodic structure of the phase domains \cite{noz84}, (1D spiral waves \cite{hag82}).
 The solution (\ref{Nozaki-Bekki}) describes a collision of the phase fronts at the borders of oppositely-oriented
 domains of the stable phase. Upon the collision both the fronts create the so called Nozaki-Bekki hole, a localized
 object which is similar to the Ising DW.

For $H=0$, following the derivation in Appendix B, we find an approximate double-front solution 
 to (\ref{simplified-GL-II}) of the form 
\begin{eqnarray}
g=\frac{1}{\sqrt{\mu}}\left({\rm e}^{k(x-x_{01})+ckt}+{\rm e}^{-k(x-x_{02})+ckt}
\right.\nonumber\\\left.
+\theta{\rm e}^{k(x-x_{01})+ckt}{\rm e}^{-k(x-x_{02})+ckt}\right),
\nonumber\\
f=\frac{1}{\theta}+{\rm e}^{k(x-x_{01})+ckt}+{\rm e}^{-k(x-x_{02})+ckt}
\nonumber\\
+\theta{\rm e}^{k(x-x_{01})+ckt}{\rm e}^{-k(x-x_{02})+ckt},
\nonumber\\
k=\sqrt{\frac{\theta}{2J}},\hspace*{3em}ck=\frac{3\theta}{2\alpha}.
\label{front_II}
\end{eqnarray}
It describes two phase fronts at the borders of similarly-oriented domains of the stable phase since, contrary to 
 the Nozaki-Bekki hole, all terms of $g$ are of the same sign. They counterpropagate into
 an unstable area. According to (\ref{front_II}), since $t\to\infty\Rightarrow g\to f/\sqrt{\mu}$, these fronts 
 annihilate upon the collision. Above I have neglected the external field because it introduces a big complexity 
 to the solution (Appendix B). However, the asymmetry of the Ginzburg-Landau function due to $H\neq 0$ 
 plays a role in the collision-induced formation of a dissipative structure since it determines the choice of the preferred 
 stable state \cite{nic83}, [though, the strength of the field is limited by the condition
 $\gamma|H|<2(\theta/3)^{3/2}/\mu^{1/2}$ ensuring that two stable homogeneous solutions to (\ref{simplified-GL})
 are of the opposite signs].

\begin{figure}
\unitlength 1mm
\begin{center}
\begin{picture}(175,29)
\put(0,-3){\resizebox{86mm}{!}{\includegraphics{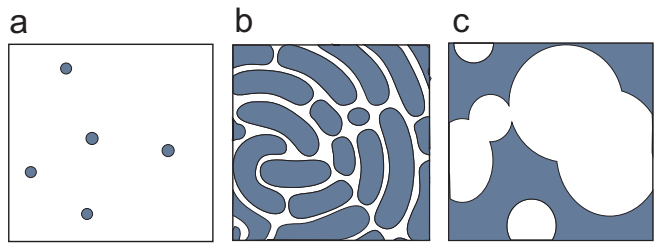}}}
\end{picture}
\end{center}
\caption{2D pattern structures in a bistable medium; (a) a bubble structure, (b) a lamellar (labyrinth) pattern,
(c) stable (white) domains propagating into (grey) unstable state. In (c) a periodic (spiral) 
structure of the domains is neglected.}
\end{figure}

\section{Discussion}

Direct observation of the front motion in ferromagnetic and ferroelectric media is difficult because relevant
 time scales are very narrow. Thus, I refer to an experiment on a (bistable) ferrocyanide-iodate-sulfite
 chemical reaction performed in a continuous-flow stirred tank reactor \cite{lee93,lee95}. When the ferrocyanide
 concentration is decreased, the morphology of the 2D chemical structure changes. At a relatively-low
 concentration, irregular stationary patterns of DWs (lamellae) appear and change into bubble structures
 with a further concentration decrease. The lamellae formation is accompanied by the property of the DW reflection
 upon their externally-driven collision (which is enforced by an intense irradiation). On increasing the critical 
 parameter (the ferrocyanide concentration), the lamellae dissapear while domains become inhomogeneous (spiral 
 structure of the domains appears) and their fronts annihilate upon the collision. I identify the transition between structures
 analogous to visualized in Figs. 1(a) and 1(b) with the decrease of the critical parameter $\beta_{1}$, and the transition 
 between structures in Figs. 1(b) and 1(c) with crossing the Bloch-Ising point $\beta_{1}=3\beta_{2}$. Thus, the predicted dynamical
 behavior of DWs of a generic (Ginzburg-Landau) model of bistable systems coincides with that observed in a chemical
 reaction below the critical point. Further increase of the ferrocyanide concentration leads to the appearance
 of wave-front reflection and self-replication of phase spots. These effects are related to differences
 in the diffusion coefficients of the chemical species; therefore, they cannot be predicted within the Ginzburg-Landau
 model. According to calculations using a two-species (Gray-Scott) model of a chemical reaction, the self-replication
 regime relates to the monostability of the system \cite{pea93}.

I conclude that the coexistence of subcritical effects of front reflection and annihilation (accompanied by 
 the creation of dissipative structures) can be explained by the appearance of domains of the unstable phase above
 the Bloch-Ising transition point.
   
\section*{Acknowledgements}

This work was supported by Polish Government Research Founds for 2010-2012 in the framework of Grant No. N N202 198039.

\appendix
\section{Derivation of DW solutions}

One manages to find an exact single-DW solution to (\ref{GL}) for $H\neq 0$ in the case ${\rm Im}(m)=0$. It 
 corresponds to an asymmetric Ising DW. In particular, inserting
\begin{eqnarray}
g_{1}=a_{1}-a_{2}{\rm e}^{2\eta},\hspace*{1.5em}f_{1}=1+{\rm e}^{2\eta},
\nonumber\\
\eta=k(x-x_{0})-\omega t,
\label{A_0}
\end{eqnarray}
into (\ref{simplified-GL-II}), one establishes the coefficients $a_{1}$, $a_{2}$, $k$, $\omega$ to be complicated
 functions of $H$. For $\gamma H\ll(\beta_{1}+\beta_{2})^{3/2}/\mu^{1/2}$, ($H$ is much weaker than the coercivity field),
 the effects of the DW asymmetry $a_{1}\neq a_{2}$ and the dependence of the DW width $1/k$ on $H$ are negligible while
 (\ref{A_0}) reproduces (\ref{DW}) and (\ref{IsingDW}). Unfortunately, an exact solution describing a Bloch DW in an external field 
 is unknown. Thus, I search for approximate solutions with ${\rm Im}(m)\neq 0$ assuming they satisfy the following requirements; 
 1) they tend to the exact (single DW or double DW) solutions of Ref. \cite{bar05} with $H\to 0$, 2) they satisfy (\ref{GL})
 at the centers of the DWs (therefore, I neglect an expected space asymmetry of ${\rm Re}(m)$, ${\rm Im}(m)$ with respect to 
 the DW center in the presence of $H\neq 0$). Within this approach, let us consider approximate single-DW solutions to 
 the bilinearized Ginzburg-Landau system (\ref{secondary_GLa}) of the form  
\begin{eqnarray}
g_{1}=a(1-{\rm e}^{2\eta})+{\rm i}b{\rm e}^{\eta},\hspace*{1.5em}f_{1}=1+{\rm e}^{2\eta},
\nonumber\\
\eta=k(x-x_{0})-\omega t,
\label{A_1}
\end{eqnarray}
Inserting (\ref{A_1}) into (\ref{secondary_GLa}), one arrives at
\begin{eqnarray}
a(\beta_{1}+\beta_{2}-\lambda)+\gamma H+{\rm e}^{\eta}{\rm i}b(\beta_{1}-\beta_{2}-\lambda
\nonumber\\
+Jk^{2}+\alpha\omega)
+{\rm e}^{2\eta}2(\gamma H-2a\alpha\omega)
\nonumber\\
+{\rm e}^{3\eta}{\rm i}b(\beta_{1}-\beta_{2}
-\lambda+Jk^{2}-\alpha\omega)
\nonumber\\
+{\rm e}^{4\eta}[-a(\beta_{1}+\beta_{2}-\lambda)+\gamma H]&=&0,
\nonumber\\
-a^{2}\mu+\lambda+{\rm e}^{2\eta}(2a^{2}\mu-b^{2}\mu+2\lambda-8Jk^{2})&&
\nonumber\\+{\rm e}^{4\eta}(-a^{2}\mu+\lambda)&=&0.
\label{A_2}
\end{eqnarray}
For $\lambda=\beta_{1}+\beta_{2}$, $a=\sqrt{\beta_{1}+\beta_{2}}/\sqrt{\mu}$, one finds the second equation 
 of (\ref{A_2}) to be satisfied, (all coefficients of the LHS expansion in ${\rm e}^{\eta}$ vanish),
 if $b=2\sqrt{\beta_{1}-3\beta_{2}}/\sqrt{\mu}$ and $|k|=\sqrt{2\beta_{2}}/\sqrt{J}$
 (the Bloch Dw) or $b=0$ and $|k|=\sqrt{\beta_{1}+\beta_{2}}/\sqrt{2J}$ (the Ising DW), whereas, for 
 $\omega=\sqrt{\mu}\gamma H/(\sqrt{\beta_{1}+\beta_{2}}\alpha)$, the LHS of the first equation of (\ref{A_2}) simplifies to
\begin{eqnarray}
\gamma H(1+{\rm e}^{\eta}{\rm i}b\frac{\sqrt{\mu}}{\sqrt{\beta_{1}+\beta_{2}}}-{\rm e}^{2\eta}2
-{\rm e}^{3\eta}{\rm i}b\frac{\sqrt{\mu}}{\sqrt{\beta_{1}+\beta_{2}}}+{\rm e}^{4\eta}).
\nonumber
\end{eqnarray}
We see that it vanishes with $H\to 0$ or at $\eta=0$ (at the DW center). Therefore, the DW described 
 by (\ref{A_1}) propagates with a constant velocity $c=\omega/k\propto H$.

Let me mention that I do not perform a systematic Hirota construction connected to inserting consecutive elements 
 of the expansion of $g_{1}$, $f_{1}$ in ${\rm e}^{\eta_{1}}$, ${\rm e}^{\eta_{2}}$ and solving a series of relevant
 equations one by one. If their solutions were not exact, it would accumulate an error in the consecutive steps. In the present 
 approach, the form of the solution (\ref{A_1}) is fixed while I simply verify the coefficients $a$, $b$, $\omega$.
 Similarly, searching for the approximate double-DW solution, I insert (\ref{doubleDW_Ia}) into (\ref{secondary_GLa}). 
 The resulting equations take the form
\begin{widetext}
\begin{eqnarray}
(1+{\rm e}^{4\eta_{1}+4\eta_{2}})[a(\beta_{1}+\beta_{2}-\lambda)+\gamma H]
+{\rm e}^{\eta_{1}}{\rm i}b\nu^{1/2}(\beta_{1}-\beta_{2}-\lambda+Jk_{1}^{2}+\alpha\omega_{1})
+{\rm e}^{2\eta_{1}}2\nu(\gamma H-2a\alpha\omega_{1})&&
\nonumber\\
+{\rm e}^{2\eta_{2}}2\nu(\gamma H
-2a\alpha\omega_{2})
+{\rm e}^{3\eta_{1}}{\rm i}b\nu^{3/2}(\beta_{1}-\beta_{2}-\lambda+Jk_{1}^{2}-\alpha\omega_{1})
+({\rm e}^{4\eta_{1}}+{\rm e}^{4\eta_{2}})\nu^{2}[-a(\beta_{1}+\beta_{2}-\lambda)+\gamma H]
&&
\nonumber\\
+{\rm e}^{2\eta_{1}+4\eta_{2}}2\nu(\gamma H+2a\alpha\omega_{1})
+{\rm e}^{4\eta_{1}+2\eta_{2}}2\nu(\gamma H+2a\alpha\omega_{2})
-{\rm e}^{\eta_{1}+2\eta_{2}}{\rm i}b\nu^{1/2}\{(1-\nu)[\beta_{1}-\beta_{2}-\lambda+Jk_{1}^{2}+4Jk_{2}^{2}&&
\nonumber\\
+\alpha\omega_{1}]
+(1+\nu)(4Jk_{1}k_{2}
+2\alpha\omega_{2})\}
+{\rm e}^{2\eta_{1}+2\eta_{2}}2[a(1-\nu^{2})(\beta_{1}+\beta_{2}-\lambda+4Jk_{1}^{2}+4Jk_{2}^{2})+
(1+\nu^{2})(8aJk_{1}k_{2}&&
\nonumber\\
+\gamma H)]
+{\rm e}^{3\eta_{1}+2\eta_{2}}{\rm i}b\nu^{1/2}\{(1-\nu)[\beta_{1}-\beta_{2}-\lambda+Jk_{1}^{2}+4Jk_{2}^{2}-\alpha\omega_{1}]
+(1+\nu)(4Jk_{1}k_{2}-2\alpha\omega_{2})\}&&
\nonumber\\
-{\rm e}^{\eta_{1}+4\eta_{2}}{\rm i}b\nu^{3/2}(\beta_{1}-\beta_{2}-\lambda+Jk_{1}^{2}+\alpha\omega_{1})
-{\rm e}^{3\eta_{1}+4\eta_{2}}{\rm i}b\nu^{1/2}(\beta_{1}-\beta_{2}-\lambda+Jk_{1}^{2}-\alpha\omega_{1})&=&0,
\nonumber\\
(1+{\rm e}^{4\eta_{1}}\nu^{2}+{\rm e}^{4\eta_{2}}\nu^{2}+{\rm e}^{4\eta_{1}+4\eta_{2}})(-a^{2}\mu+\lambda)
+({\rm e}^{2\eta_{1}}+{\rm e}^{2\eta_{1}+4\eta_{2}})\nu(2a^{2}\mu+2\lambda-8Jk_{1}^{2}-b^{2}\mu)
+{\rm e}^{2\eta_{1}+2\eta_{2}}2&&
\nonumber\\
\times[-(1-\nu^{2})8Jk_{1}k_{2}+(1+\nu^{2})(-a^{2}\mu+\lambda-4Jk_{1}^{2}-4Jk_{2}^{2})+\nu b^{2}\mu]
+({\rm e}^{2\eta_{2}}+{\rm e}^{4\eta_{1}+2\eta_{2}})\nu(2a^{2}\mu+2\lambda-8Jk_{2}^{2})
&=&0.
\label{A_4}
\end{eqnarray}
\end{widetext}
The second equation of (\ref{A_4}) is fulfilled for $\lambda$, $a$, $b$ determined above and for 
 $k_{1}$, $k_{2}$, $\nu$ of (\ref{doubleDW_II}). When $H=0$, the first equation of (\ref{A_4})
 is satisfied for this sort of parameters according to \cite{bar05}. 
 In the vicinity of the DW centers in a state of well separated DWs; the regimes 
 $\eta_{1}+{\rm ln}(\nu)/2\approx 0$, ${\rm e}^{\eta_{2}+{\rm ln}(\nu)/2}\approx 0$
 or $\eta_{2}+{\rm ln}(\nu)/2\approx 0$, ${\rm e}^{\eta_{1}+{\rm ln}(\nu)/2}\approx 0$, the system (\ref{A_4}) 
 reproduces (\ref{A_2}); therefore, the centers of both the DWs counterpropagate with constant velocities.
 
\section{Derivation of double-front solution}

Looking for the approximate double-front solution (\ref{front_II}), I define
 an error function as the absolute value of the LHS of (\ref{simplified-GL-II}) for $H=0$
\begin{eqnarray}
{\rm Err}(x,t)\equiv|f[(-\alpha D_{t}+JD_{x}^{2})g\cdot f]
\nonumber\\
+g\left[(-JD_{x}^{2}+\theta)f\cdot f-\mu\theta g^{2}\right]|.
\label{Err}
\end{eqnarray}
Inserting
\begin{eqnarray}
g=h\left({\rm e}^{\eta_{1}}+{\rm e}^{\eta_{2}}
+\varphi{\rm e}^{\eta_{1}+\eta_{2}}\right),
\nonumber\\
f=\varphi^{-1}+{\rm e}^{\eta_{1}}+{\rm e}^{\eta_{2}}
+\varphi{\rm e}^{\eta_{1}+\eta_{2}},
\end{eqnarray}
where $\eta_{1}\equiv k(x-x_{01})+ckt$, $\eta_{2}\equiv-k(x-x_{02})+ckt$, into (\ref{Err}), 
 we take $\varphi=\theta$ in order to ensure faster divergence of the front centers 
 in the limit $\theta\to 0^{+}$ than the divergence of the front widths in this limit. 
 We arrive at
\begin{eqnarray}
{\rm Err}(x,t)&=&|[{\rm e}^{3\eta_{1}+3\eta_{2}}\theta^{4}+({\rm e}^{2\eta_{1}+3\eta_{2}}+{\rm e}^{3\eta_{1}+2\eta_{2}})3\theta^{3}
\nonumber\\
&&+({\rm e}^{3\eta_{1}+\eta_{2}}+{\rm e}^{\eta_{1}+3\eta_{2}})3\theta^{2}+({\rm e}^{3\eta_{1}}+{\rm e}^{3\eta_{2}})\theta]
\nonumber\\
&&\times b(1-\mu b^{2})+({\rm e}^{2\eta_{1}}+{\rm e}^{2\eta_{2}})b(2-\alpha ck/\theta
\nonumber\\
&&-Jk^{2}/\theta)+{\rm e}^{2\eta_{1}+2\eta_{2}}\theta^{2}b(8-6\mu b^{2}-2\alpha ck/\theta)
\nonumber\\
&&+({\rm e}^{\eta_{1}+2\eta_{2}}+{\rm e}^{2\eta_{1}+\eta_{2}})\theta b(7-3\mu b^{2}
-3\alpha ck/\theta
\nonumber\\
&&+Jk^{2}/\theta)+{\rm e}^{\eta_{1}+\eta_{2}}b(5-4\alpha ck/\theta+6Jk^{2}/\theta)
\nonumber\\
&&+({\rm e}^{\eta_{1}}+{\rm e}^{\eta_{2}})\theta^{-1}b(1-\alpha ck/\theta+Jk^{2}/\theta)|.
\label{B_1}
\end{eqnarray}
For $h=1/\sqrt{\mu}$, $k=\sqrt{\theta/2J}$, $ck=3\theta/(2\alpha)$, all the coordinate $x$ dependent 
 terms of (\ref{B_1}) vanish, leading to
\begin{eqnarray}
{\rm Err}(x,t)={\rm Err}(t)\sim|{\rm e}^{2\eta_{1}+2\eta_{2}}\theta^{2}-2{\rm e}^{\eta_{1}+\eta_{2}}|
\nonumber\\
=|{\rm e}^{4ckt-2k(x_{01}-x_{02})}\theta^{2}-2{\rm e}^{2ckt-k(x_{01}-x_{02})}|.
\end{eqnarray}
Before the collision, when the fronts are well separated, Eq. (\ref{simplified-GL-II}) [equivalent to
 ${\rm Err}(x,t)=0$] is approximately fulfilled in the vicinity of the front centers [the regimes
 $\eta_{1}+{\rm ln}(\theta)\approx 0$, ${\rm e}^{\eta_{2}+{\rm ln}(\theta)}\approx 0$ or 
 $\eta_{2}+{\rm ln}(\theta)\approx 0$, ${\rm e}^{\eta_{1}+{\rm ln}(\theta)}\approx 0$] as well as
 in the area of unstable phase separating the fronts (${\rm e}^{\eta_{1}+{\rm ln}(\theta)}\approx 0$, 
 ${\rm e}^{\eta_{2}+{\rm ln}(\theta)}\approx 0$). Then 
\begin{eqnarray}
m^{'}&=&\frac{1}{\sqrt{\mu}}\frac{{\rm e}^{\eta_{1}}+{\rm e}^{\eta_{2}}+\theta{\rm e}^{\eta_{1}+\eta_{2}}}{
1/\theta+{\rm e}^{\eta_{1}}+{\rm e}^{\eta_{2}}+\theta{\rm e}^{\eta_{1}+\eta_{2}}}\\
&=&\left\{\begin{array}{cc}
\frac{1}{\sqrt{\mu}}\frac{{\rm e}^{\eta_{1}+{\rm ln}(\theta)}}{1+{\rm e}^{\eta_{1}+{\rm ln}(\theta)}} &\hspace*{0.5em}\eta_{1}\sim-{\rm ln}(\theta),
{\rm e}^{\eta_{2}+{\rm ln}(\theta)}\sim 0,\\
\frac{1}{\sqrt{\mu}}\frac{{\rm e}^{\eta_{2}+{\rm ln}(\theta)}}{1+{\rm e}^{\eta_{2}+{\rm ln}(\theta)}} &\hspace*{0.5em}\eta_{2}\sim-{\rm ln}(\theta),
{\rm e}^{\eta_{1}+{\rm ln}(\theta)}\sim 0.
\end{array}
\right.\nonumber
\end{eqnarray}
Assuming the motion to start at $t=0$, and choosing the reference frame such that $x_{02}=-x_{01}$,
 the front centers are initially at the points $x=\pm x_{1}=\pm(x_{01}-{\rm ln}(\theta )\sqrt{2J/\theta})$
 and they meet at the point $x=0$ after the time $t^{'}=x_{1}/c$. Let us notice that up to the time 
 $t^{''}=t^{'}+\frac{\rm ln(2)}{2ck}$, the function ${\rm Err}(t)$ is a decreasing one and ${\rm Err}(t^{''})=0$. 
 We conclude that the approximate solution (\ref{front_II}) satisfies well (\ref{simplified-GL-II}) 
 at the beginning of the motion (at $t=0$) and the accuracy of approximating the solution by the functions (\ref{front_II})
 increases with time up to $t=t^{''}$ which corresponds to the moment of finishing the front collision (the moment when the fronts 
 of the width $1/k$ finish their passing through each other with the relative velocity $2c$). 

The application of a similar analysis to the field induced collision of DWs below the Bloch-Ising 
 transition, studying (\ref{secondary_GLa}) separately from (\ref{secondary_GLb}), does not work. 
 Utilizing the argumentation (of Sec. II) based on the time-reversal symmetry is an alternative.

\end{document}